\documentclass[aps,pra,reprint, amsmath, amssymb,superscriptaddress,nofootinbib,crop=false]{revtex4-1}

\usepackage[subpreambles=true]{standalone}
\usepackage{import}
\usepackage{blindtext}

\usepackage{bm}
\usepackage[retainorgcmds]{IEEEtrantools}
\usepackage{graphicx}
\usepackage{mathrsfs}
\usepackage{amsmath}
\usepackage{amssymb}
\usepackage{color}
\usepackage{amsfonts}
\usepackage{times,txfonts}
\usepackage{nicefrac}
\usepackage[colorlinks=true,linkcolor=blue,urlcolor=blue,citecolor=blue,pdfusetitle]{hyperref}
\usepackage{amsmath}

\newcommand{\tr}{\text{tr}}

\usepackage{subfiles} 

\begin{document}

\title{Thermodynamic uncertainty relation for quantum entropy production}
\date{\today}
\author{Domingos S. P. Salazar}
\affiliation{Unidade de Educa\c c\~ao a Dist\^ancia e Tecnologia,
Universidade Federal Rural de Pernambuco,
52171-900 Recife, Pernambuco, Brazil}

\begin{abstract}
In quantum thermodynamics, entropy production is usually defined in terms of the quantum relative entropy between two states. We derive a lower bound for the quantum entropy production in terms of the mean and variance of quantum observables, which we will refer to as a thermodynamic uncertainty relation (TUR) for the entropy production. In the absence of coherence between the states, our result reproduces classic TURs in stochastic thermodynamics. For the derivation of the TUR, we introduce a lower bound for a quantum generalization of the $\chi^2$ divergence between two states and discuss its implications for stochastic and quantum thermodynamics, as well as the limiting case where it reproduces the quantum Cramér-Rao inequality.

\end{abstract}
\maketitle{}

{\bf \emph{Introduction -}} 
Entropy flows between systems, but it can also be produced. Entropy production is a signature of nonequilibrium and a core concept in thermodynamics. In stochastic thermodynamics \cite{LandiReview2021,Seifert2012}, physical quantities are defined at the trajectory level, including the (classical) entropy production $\Sigma_{\text{cl}}$, which is a random variable that satisfies the second law of thermodynamics when averaged over trajectories.
\begin{equation}
\label{secondlaw}
    \langle \Sigma_{\text{cl}} \rangle \geq 0.
\end{equation}
Although the second law is always satisfied, there are situations when one has information about an observable. For instance, the mean $\langle \phi \rangle$ and variance of a current $\langle \langle \phi \rangle \rangle$. In this case, there are a series of results called thermodynamic uncertainty relations (TURs) that set lower bounds for $\langle \Sigma_{\text{cl}} \rangle$, thereby improving upon the second law.
\begin{equation}
\label{TUR}
\langle \Sigma_{\text{cl}} \rangle \geq f_{cl}(\langle \phi \rangle, \langle \langle \phi \rangle \rangle)\geq 0.
\end{equation}
There are several versions of TURs \cite{VoV2020,Erker2017,MacIeszczak2018,Carollo2019,Goold2014b}, which may involve more than one variance in the argument of (\ref{TUR}). In addition to the original formulation for steady states \cite{Barato2015,Gingrich2016,Horowitz2019}, we also have, for instance, the exchange TUR \cite{Timpanaro2019B,Hasegawa2019} and the hysteretic TUR \cite{Potts2019,Proesmans2019,Gianluca2022,Salazar2022d}.

In quantum thermodynamics \cite{Campisi2011,Batalhao2014,Hanggi2015,Esposito2009}, the usual setup \cite{LandiReview2021} consists of a system (S) and an environment (E), with the initial state given by $\rho_{SE} = \rho_S \otimes \rho_E$ prepared independently, although the treatment that follows also allows for initial quantum correlations. The whole system (S+E) evolves with a unitary $\mathcal{U}$, such that $\rho'_{SE} = \mathcal{U} \rho_{SE} \mathcal{U}^\dagger$ is the final state. In this setup, the quantum entropy production is defined as
\begin{equation}
\label{QEP}
\Sigma := S(\rho'_{SE}||\rho_S' \otimes \rho_E),
\end{equation}
where $\rho_S':=\tr_E (\rho_{SE}')$ is the final reduced state of the system after tracing over the environment and $S(\rho||\sigma) = \tr(\rho \ln \rho) - \tr(\rho \ln \sigma)$ is the quantum relative entropy. There are few advantages for definition (\ref{QEP}), whose protocol is known as bath reset. Other definitions, such as both reset, $S(\rho_{SE}'||\rho_S \otimes \rho_E)$, are also contemplated in this analysis without loss of generality. For the sake of simplicity, we consider definition (\ref{QEP}). Not surprisingly, there are also quantum TURs for quantum systems \cite{Liu2019,VanVu2023,Miller2021,Pires2021,Guarnieri2019,Hasegawa2019a,VuT2020,VanVu2019,VanVu2021,Hasegawa2020b}, typically involving different measures of divergence other than the entropy production itself, such as survival activity \cite{Hasegawa2021}, which has been verified in quantum computing \cite{Ishida2024}.

In this letter, we investigate if a lower bound for the quantum entropy production (\ref{QEP}) exists in terms of means and variances of observables. In this case, we consider any observable $\hat{\theta}$ acting on system + environment (S+E). We obtain the following result,
\begin{equation}
\label{main}
\Sigma \geq F(\langle \hat{\theta} \rangle_\rho - \langle \hat{\theta} \rangle_\sigma, \langle \langle \hat{\theta} \rangle\rangle_\rho, \langle \langle \hat{\theta} \rangle\rangle_\sigma),
\end{equation}
where $\rho=\rho_{SE}'$ and $\sigma=\rho_S' \otimes \rho_E$ (or any other protocol for the backward state), with notation $\langle \hat{\theta}\rangle_\rho := \tr(\hat{\theta}\rho)$ and $\langle \langle \hat{\theta} \rangle \rangle_\rho := \langle (\hat{\theta}-\langle \hat{\theta} \rangle_\rho\rangle_\rho$, and $F(x,y,z):=\int_0^1 \frac{\lambda x^2}{(1-\lambda)y+\lambda z + (1-\lambda)\lambda x^2}d\lambda $, with explicit form discussed later in the paper. 

In relation (\ref{main}), the right-hand side depends only on the mean and variance of any observable. In fact, in the absence of coherence ($[\rho,\sigma]=[\rho,\hat{\theta}]=0$), (\ref{main}) reproduces a thermodynamic uncertainty relation in stochastic thermodynamics \cite{VoV2020} and a result in information theory \cite{Nishiyama2020}. For this reason, we refer to (\ref{main}) as a thermodynamic uncertainty relation for quantum entropy production.

Furthermore, we obtained our result (\ref{main}) from a theorem that represents a quantum equivalent of the variation representation of the $\chi^2$ divergence or the Hammersley–Chapman–Robbins bound. First, we define a generalization of the quantum $\chi^2$ divergence between states $\rho$ and $\sigma$ as follows,
\begin{equation}
\label{qchi}
\chi^2_\lambda[\rho,\sigma]:=\lambda^2\sum_{ij}\frac{(p_i-q_j)^2}{(1-\lambda)p_i + \lambda q_j}|\langle p_i |q_j \rangle|^2,
\end{equation}
where we considered the spectral decomposition $\rho=\sum_i p_i |p_i\rangle \langle p_i|$, $\sigma=\sum_j q_j |q_j\rangle \langle q_j|$
and $\lambda \in [0,1]$. Now we have our main result,
\begin{equation}
\label{secondmain}
\chi^2_\lambda[\rho,\sigma] \geq \frac{\lambda^2 (\langle \hat{\theta}\rangle_\rho - \langle \hat{\theta}\rangle_\sigma)^2}{\lambda \langle\langle \hat{\theta}\rangle \rangle_\rho+(1-\lambda)\langle \langle \hat{\theta}\rangle\rangle_\sigma + \lambda (1-\lambda)(\langle \hat{\theta}\rangle_\rho - \langle \hat{\theta}\rangle_\sigma)^2}.
\end{equation}
The characterization of $\chi^2_\lambda[\rho,\sigma]$ and its relation to the Fisher information is discussed later. The lower bound (\ref{main}) is a consequence of (\ref{secondmain}), as shown in the formalism section. However, (\ref{secondmain}) has other applications, as discussed below.

This letter is organized as follows. First, we prove (\ref{secondmain}) and the thermodynamic uncertainty relation for quantum entropy production (\ref{main}). Then, we discuss the applications of (\ref{secondmain}) in various scenarios, including limiting cases where it reproduces results in stochastic thermodynamics and the quantum Cramér-Rao inequality.

{\bf \emph{Formalism -}} In terms of any distributions $P,Q$ in a countable set $\mathcal{S}$, we have the $\chi^2$ divergence as the usual expression, $\chi^2(P|Q) := \sum_{s \in \mathcal{S}} (P_{s}-Q_{s})^2/Q_{s}$, and now we define the following (still classic) generalization,
\begin{equation}
\label{form1}
\chi^2_\lambda (P|Q) := \chi^2(P|(1-\lambda)P+\lambda Q)=\sum_s \frac{\lambda^2(P_s - Q_s)^2}{(1-\lambda)P(s)+\lambda Q(s)},
\end{equation}
for $\lambda \in [0,1]$, such that $\chi^2_1(P|Q)=\chi^2(P|Q)$. Now we explore a result known as the variation representation of $\chi^2$ divergence, which is also intimately related to the Hammersley–Chapman–Robbins bound \cite{Hammersley1950}. Let $\Theta$ be a complex random variable, we have
\begin{equation}
\label{form2}
\chi^2(P|Q) \geq \frac{|\langle \Theta \rangle_P - \langle \Theta \rangle_Q|^2 }{\langle \langle \Theta \rangle \rangle_{Q}},
\end{equation}
with notation $\langle \Theta \rangle_p = \sum_s p_s \Theta_s$ and $\langle \langle \Theta \rangle \rangle_p = \langle |\Theta - \langle \Theta \rangle_p |^2\rangle_p$. Although the original relation is for real variables, we provide a simple proof of (\ref{form2}) for complex variables (see Appendix). We apply (\ref{form2}) for the pair of probabilities $(P,(1-\lambda)P + \lambda Q)$ using definition (\ref{form1}), resulting in 
\begin{equation}
\label{form3}
\chi^2_\lambda (P|Q) \geq \frac{\lambda^2 |\langle \Theta\rangle_P - \langle \Theta\rangle_Q|^2}{\lambda \langle\langle \Theta\rangle \rangle_P+(1-\lambda)\langle \langle \Theta\rangle\rangle_Q + \lambda (1-\lambda)|\langle \Theta\rangle_P - \langle \Theta\rangle_Q|^2},
\end{equation}
which is the classic version of our result (\ref{secondmain}) for complex random variables.

For the quantum version of (\ref{form3}), we use a successful a strategy known as Nussbaum-Szkoła (NS) distributions \cite{Nussbaum2009,Androulakis2023a,Salazar2023d,Androulakis2024} that maps $n$ dimensional states with spectral decomposition $\rho = \sum_i p_i|p_i\rangle \langle p_i|$ and $\sigma=\sum_j q_j |q_j\rangle \langle q_j|$ into the following $n^2$ dimensional classic distributions: $P_{ij}:=p_i|\langle p_i|q_j \rangle |^2$ and $Q_{ij}:=q_j |\langle p_i | q_j \rangle|^2$. In the treatment that follows, we may have $n=\mathbb{N}$. Check that $\sum_{ij}P_{ij}= \sum_{ij}Q_{ij}=1$ and $P_{ij}\geq 0$, $Q_{ij}\geq 0$, for all $i,j$. Using the same idea, we map any Hermitian operator in the following complex random variable in $n^2$ dimensions,
\begin{equation}
\label{form3b}
\Theta_{ij} := \frac{\langle p_i | \hat{\theta} | q_j\rangle}{\langle p_i|q_j\rangle},
\end{equation}
for $\langle p_i | q_j \rangle \neq 0$, and $\Theta_{ij}:=0$, for $\langle p_i | q_j\rangle = 0$. 
Applying (\ref{form1}) for the NS distributions ($P_{ij},Q_{ij}$), we obtain
\begin{equation}
\label{form4}
\chi^2_\lambda(P|Q) = \lambda^2\sum_{ij}\frac{(p_i-q_j)^2}{(1-\lambda)p_i + \lambda q_j}|\langle p_i |q_j \rangle|^2 = \chi^2_\lambda[\rho,\sigma],
\end{equation}
as defined in (\ref{qchi}). Following previous treatments \cite{Salazar2023d}, we also have from the definition (\ref{form4}), 
\begin{equation}
\label{form5}
\langle \Theta \rangle_P = \sum_{ij, \langle p_i | q_j \rangle \neq 0}
\frac{\langle p_i | \hat{\theta} |q_j \rangle}{\langle p_i | q_j \rangle} |\langle p_i | q_j \rangle|^2 = \langle \hat{\theta} \rangle_\rho,
\end{equation}
and $\langle \Theta \rangle_Q = \langle \hat{\theta}\rangle_\sigma$. Similarly, we also have $\langle \hat{\theta}^2 \rangle_\rho =\sum_{ij} p_i |\langle p_i |\hat{\theta}|q_j \rangle|^2 \geq \sum_{ij;\langle q_j|p_i\rangle \neq 0} p_i |\langle p_i |\hat{\theta}|q_j \rangle|^2 = \sum_{ij} P_{ij}|\Theta_{ij}|^2 = \langle |\Theta|^2 \rangle_P$,
which, when combined with (\ref{form5}), it results in
\begin{equation}
\label{form6}
\langle \langle \hat{\theta} \rangle \rangle_\rho \geq \langle \langle \Theta \rangle \rangle_P,
\end{equation}
and $\langle \langle \hat{\theta} \rangle \rangle_\sigma \geq \langle \langle \Theta \rangle \rangle_Q$. Defining the real function
\begin{equation}
\label{form7}
f_\lambda(x,y,z):=\frac{\lambda x^2}{(1-\lambda)y+\lambda z + (1-\lambda)\lambda x^2},
\end{equation}
for $\lambda \in [0,1]$, we obtain the following inequality immediately from (\ref{form5}) and (\ref{form6}), $f_\lambda(|\langle \Theta \rangle_P - \langle \Theta \rangle_Q|, \langle \langle \Theta \rangle \rangle_P, \langle \langle \Theta \rangle \rangle_Q) \geq f_\lambda(\langle \hat{\theta} \rangle_\rho - \langle \hat{\theta}\rangle_\sigma ,\langle \langle \hat{\theta}\rangle\rangle_\rho , \langle \langle \hat{\theta}\rangle\rangle_\sigma)$, because $f_\lambda (x,y,z)$ is decreasing in $y\geq0$ and $z\geq0$. Finally, from (\ref{form4}) and (\ref{form3}), we have
\begin{eqnarray}
\label{form8}
\chi^2_\lambda(\rho,\sigma)=\chi^2_\lambda(P|Q) \geq \lambda f_\lambda(|\langle \Theta \rangle_P - \langle \Theta \rangle_Q|, \langle \langle \Theta \rangle \rangle_P, \langle \langle \Theta \rangle \rangle_Q) 
\notag
\\
\geq \lambda f_\lambda(\langle \hat{\theta} \rangle_\rho - \langle \hat{\theta}\rangle_\sigma ,\langle \langle \hat{\theta}\rangle\rangle_\rho , \langle \langle \hat{\theta}\rangle\rangle_\sigma),
\end{eqnarray}
which is our result (\ref{secondmain}). The application to quantum thermodynamics (\ref{main}) goes as follows. First, check that the Kullback-Leibler divergence between the NS distributions $P$ and $Q$ results in the quantum relative entropy between the underlying states $\rho$ and $\sigma$ \cite{Salazar2023d,Salazar2024b},
\begin{equation}
\label{form9}
D(P|Q):=\sum_{ij}P_{ij}\ln (P_{ij}/Q_{ij}) =S(\rho||\sigma).
\end{equation}
Then, we use the following integral representation of the KL divergence \cite{Nishiyama2020}, 
\begin{equation}
\label{form10}
S(\rho||\sigma)=D(P|Q) = \int_0^1 \chi^2_\lambda (P|Q)\frac{d\lambda}{\lambda},
\end{equation}
and combine it with (\ref{form8}), resulting in
\begin{equation}
\label{form11}
S(\rho||\sigma) \geq \int_0^1 f_\lambda(\langle \hat{\theta} \rangle_\rho - \langle \hat{\theta}\rangle_\sigma ,\langle \langle \hat{\theta}\rangle\rangle_\rho , \langle \langle \hat{\theta}\rangle\rangle_\sigma) d\lambda,
\end{equation}
which is our result (\ref{main}). For completeness, the explicit form of $F(x,y,z)=\int_0^1 f_\lambda(x,y,z)d\lambda$ used in (\ref{form11}) is given in \cite{Nishiyama2020}, so that the lower bound (\ref{form11}) reads
\begin{equation}
\label{form12}
S(\rho||\sigma) \geq r\ln(r/s) + (1-r)\ln\big(\frac{1-r}{1-s}\big),
\end{equation}
for $r,s \in [0,1]$, where $r:=1/2+b/(4av)$, $s:=r-a/2v$, $a:=\langle \hat{\theta} \rangle_\rho - \langle \hat{\theta} \rangle_\sigma$, $b:=a^2+\langle \langle \hat{\theta}\rangle \rangle_\sigma - \langle \langle \hat{\theta} \rangle \rangle_\rho$, $v:=\sqrt{\langle \langle \hat{\theta} \rangle \rangle_\rho +b^2/(4a^2)}$. Note that the rhs in (\ref{form12}) depends only on the mean and variance of $\hat{\theta}$ with respect to $\sigma$ and $\rho$. Choosing $\rho:=\rho_{SE}'$ and $\sigma:=\rho_S' \otimes \rho_E$ (or any other protocol for $\sigma$), we obtain $S(\rho||\sigma)=\Sigma$, which makes (\ref{form12}) a thermodynamic uncertainty relation for quantum entropy production (\ref{main}). Previously, the lower bound for the quantum relative entropy \cite{Salazar2023d} relied on the definition of a its dual $\Sigma^*=\Sigma(\rho_S'\otimes \sigma_E || \rho_{SE}')$, but (\ref{main}) poses a lower bound for $\Sigma$ without the introduction of the dual. Inequality (\ref{form12}) is essentially a quantum generalization of a result in stochastic thermodynamics \cite{VoV2020} and information theory \cite{Nishiyama2020} as discussed below.

{\bf \emph{Discussion-}} In the following sections, we discuss aspects of our main result (\ref{secondmain}) and implications in stochastic and quantum thermodynamics, as well as a proper characterization of $\chi_\lambda^2(\rho,\sigma)$ and particular cases.

{\bf \emph{a. Stochastic thermodynamics -}} Note that the particular case of (\ref{form12}) in the absence of coherence, $[\rho,\sigma]=[\rho,\hat{\theta}]=0$, with $\rho$ and $\sigma$ written as probabilities $p=(p_1,...,p_n)$, $q=(q_1,...,q_n)$, and the random variable $\theta=(\theta_1,...,\theta_n)$ such that $p=P_F(\Gamma)$ and $q=P_B(\Gamma^\dagger)$, where $\Gamma$ represents a trajectory in stochastic thermodynamics and $\Gamma^\dagger$ is the inverse trajectory, $(\Gamma^\dagger)^\dagger = \Gamma$. The observable is $\theta(\Gamma)$ and we obtain the following result in information theory \cite{Nishiyama2020} from (\ref{main}),
\begin{equation}
\label{discussion1}
\langle \Sigma_{cl} \rangle_p \geq F(\langle \theta \rangle_p - \langle \theta \rangle_q, \langle \langle \theta \rangle \rangle_p, \langle \langle \theta \rangle \rangle_q),
\end{equation}
where $\langle \Sigma_{\text{cl}}\rangle_p:=\sum_\Gamma P_F(\Gamma)(\ln (P_F(\Gamma)/P_B(\Gamma^\dagger)$ is the entropy production. Moreover, if $\theta(\Gamma)$ is an observable with property $\theta(\Gamma^\dagger)=\epsilon\theta(\Gamma)$, $\epsilon \in \{-1,1\}$, then (\ref{discussion1}) reproduces a lower bound for the entropy production in stochastic thermodynamics \cite{VoV2020}. Furthermore, as already pointed out in \cite{VoV2020}, expression (\ref{discussion1}) for $P_F=P_B$ reproduces the thermodynamic uncertainty relation from the exchange fluctuation theorem \cite{Hasegawa2019,Timpanaro2019B},
\begin{equation}
\label{xTUR}
\frac{\langle \langle \theta \rangle \rangle_p}{\langle \theta \rangle_p^2} \geq [\sinh(g(\langle \Sigma_{cl} \rangle_p)/2)]^{-2},
\end{equation}
where $g(x)$ is the inverse of $h(x):=x\tanh(x/2)$ for $x>0$. 

{\bf \emph{b. Quantum systems -}} Now that we established a connection with stochastic thermodynamics, we analyse the implications of our results in quantum information and thermodynamics. First, we write an alternative characterization of $\chi_\lambda^2(\rho,\sigma)$ from (\ref{qchi}) for spectral decomposition $\rho=\sum_i p_i |p_i\rangle \langle p_i|$, $\sigma=\sum_j q_j |q_j\rangle \langle q_j|$, given by
\begin{equation}
\label{qchioperator}
\chi_\lambda^2(\rho,\sigma)=\tr\big[\big(|\Psi_\rho\rangle \langle \Psi_\rho| - |\Psi_\sigma\rangle \langle \Psi_\sigma|\big)\Omega_\lambda(\rho,\sigma)\big],
\end{equation}
with $|\Psi_\rho\rangle:=\sum_i \sqrt{p_i}|p_i, p_i\rangle$, $|\Psi_\sigma \sum_j \rangle:=\sqrt{q_j}|q_j, q_j\rangle$ and $\Omega_\lambda (\rho,\sigma):=\lambda^2[\rho\otimes I - I\otimes\sigma][(1-\lambda)\rho \otimes I + I \otimes \lambda\sigma]^{-1}$, where, when $[(1-\lambda)\rho \otimes I + I \otimes \lambda\sigma]$ is not full rank, the inverse is interpreted as the Moore-Penrose
pseudoinverse. Note that $\Omega_\lambda(\rho,\sigma)|p_i,q_j\rangle=\lambda^2(p_i-q_j)/((1-\lambda)p_i + \lambda q_j)$, $|\langle \Psi_\rho|p_i,q_j\rangle|^2 = p_i |\langle p_i | q_j \rangle|^2$ and $|\langle \Psi_\rho|p_i,q_j\rangle|^2 = q_j |\langle p_i | q_j \rangle|^2$, so that expressions (\ref{qchi}) and (\ref{qchioperator}) coincide. Thus we have from (\ref{qchioperator}), $\chi_\lambda^2(\rho,\sigma)=\langle \Psi_\rho | \Omega_\lambda (\rho,\sigma) |\Psi_\rho\rangle - \langle \Psi_\sigma | \Omega_\lambda (\rho,\sigma) |\Psi_\sigma\rangle$, which is an alternative form of defining $\chi^2_\lambda(\rho,\sigma)$.

For the specific case $\lambda=1/2$, we have a quantum analog of the triangular discrimination,
\begin{equation}
\label{td}
    \chi_{1/2}^2(\rho,\sigma)=\frac{1}{2}\sum_{ij} \frac{(p_i-q_j)^2}{p_i+q_j} |\langle p_i | q_j \rangle|^2:=\delta(\rho,\sigma),
\end{equation}
and, when combined with (\ref{secondmain}) and after some manipulation, it results in the following quantum thermodynamic uncertainty relation for such generalized triangular discrimination,
\begin{equation}
\label{tdTUR}
\frac{\langle \langle \hat{\theta} \rangle \rangle_\rho + \langle \langle \hat{\theta} \rangle \rangle_\sigma}{(1/2)(\langle \hat{\theta}\rangle_\rho - \langle \hat{\theta}\rangle_\sigma)^2} \geq \frac{1-\delta(\rho,\sigma)}{\delta(\rho,\sigma)},
\end{equation}
for $\langle \hat{\theta}\rangle_\rho \neq \langle \hat{\theta}\rangle_\sigma$, which is few steps away from the uncertainty relation for the symmetric Petz-Rényi relative entropies \cite{Salazar2024b}, which includes the symmetric quantum relative entropy \cite{Salazar2023d}. The operator representation of $\delta(\rho,\sigma)$ is given by (\ref{qchioperator}) for $\lambda=1/2$, so that $\Omega_{1/2} (\rho,\sigma):=(1/2)[\rho\otimes I - I\otimes\sigma][\rho \otimes I + I \otimes\sigma]^{-1}$. Therefore, our result (\ref{secondmain}) is also connected to a family of symmetric quantum uncertainty relations (which also include TURs in stochastic thermodynamics, such as the hysteretic TUR, as a particular case). One could think of (\ref{secondmain}), and the particular case (\ref{tdTUR}) as well as (\ref{form12}) in terms of quantum channels, where $\mathcal{E}_t$ is a completely positive trace preserving (CPTP) map and $\rho=\rho(t)=\mathcal{E}_t(\sigma)$ and $\sigma=\rho(0)$. 

{\bf \emph{c. Unitary dynamics - }} Let us take our main result (\ref{secondmain}) and assume that $\sigma = U\rho U^\dagger$, where $U$ is a unitary operator. This case is interesting, because it might represent the time evolution of a quantum system. We have the spectral decomposition of $\rho=\sum_i p_i|p_i\rangle\langle p_i|$, which makes $\sigma = \sum_j p_j U|p_i\rangle\langle p_j|U^\dagger$. In this case, we have a specific form for $\chi_\lambda^2(\rho,U\rho U^\dagger)$ from (\ref{qchi}) given by
\begin{equation}
\label{unit1}
\chi_\lambda^2(\rho,U\rho U^\dagger) = \lambda^2\sum_{ij}\frac{(p_i-p_j)^2}{(1-\lambda)p_i + \lambda p_j}|\langle p_i |U| p_j\rangle|^2,
\end{equation}
and the resulting TUR (\ref{secondmain}) with (\ref{unit1}) depends only on $\rho$ and $U$. Particularly, if one takes $\lambda=1/2$ and $U:=\exp(-iH\tau)$, $H^\dagger = H$, using (\ref{td}), denoting $\rho=\rho(t)$ and $\sigma=\rho(t+\tau)=U \rho(t) U^\dagger$, we have the following expansion,
\begin{equation}
\label{unit2}
\chi_{1/2}^2(\rho,U\rho(t) U^\dagger) = \delta(\rho,\sigma)= F_Q [\rho(t),H] \frac{\tau^2}{4} + \mathcal{O}(\tau^4),
\end{equation}
where the first term in the expansion contains the quantum Fisher information of $\rho$ with respect to observable $H$, obtained from the unitary perturbation $\rho_{t+\tau}=U\rho_t U^\dagger$,
\begin{equation}
\label{unit3}
F_Q [\rho,H]:=2\sum_{ij} \frac{(p_i-p_j)^2}{p_i + p_j}|\langle p_i |H| p_j\rangle|^2.
\end{equation}
Finally, using (\ref{tdTUR}) and (\ref{unit2}), we obtain
\begin{equation}
\label{unit4}
\frac{4\langle \langle \hat{\theta} \rangle \rangle_{t}}{{(\langle \hat{\theta}\rangle_{t+\tau}} -\langle \hat{\theta}\rangle_t)^2/\tau^2} \geq \frac{4}{F_Q[\rho(t),H]} + \mathcal{O}(\tau^2),
\end{equation}
and taking the limit $\tau \rightarrow 0$ in (\ref{unit4}) results in the quantum Cramér-Rao inequality,
\begin{equation}
\label{unit5}
\frac{\langle \langle \hat{\theta} \rangle \rangle_{t}}{(\partial_t\langle \hat{\theta}\rangle_t)^2} \geq \frac{1}{F_Q[\rho(t),H]}.
\end{equation}
The quantum CR inequality has found applications in quantum thermodynamics in the form of uncertainty relations \cite{Hasegawa2021,Ishida2024}. Although the inequality (\ref{unit5}) works for any real parameter $t$ disturbing the state $\rho(t)$, it is interesting to understand (\ref{unit5}) with respect to time ($t$) and a Hamiltonian ($H$). In the light of this result (\ref{unit5}), one could interpret (\ref{secondmain}) as a finite element version (instead of differential) of the quantum Cramér-Rao inequality.

{\bf \emph{d. Quantum thermodynamics - }} We turn our focus to the setup of quantum thermodynamics described in the introduction (\ref{QEP}), where $\Sigma=S(\rho||\sigma)$ defines the quantum entropy production, with $\rho=\rho_{SE}'=\mathcal{U}(\rho_S \otimes \rho_E) \mathcal{U}^\dagger$ and $\sigma = \rho_S' \otimes \rho_E$. As a particular application of (\ref{main}), consider the operator $\hat{\theta}=I_S \otimes \ln\rho_E$, such that $\langle \hat{\theta} \rangle_{\rho} = \tr(\rho_E'\ln \rho_E)$ and $\langle \hat{\theta} \rangle_{\sigma} = \tr(\rho_E\ln \rho_E)$. Now we define the entropy flux as  $\Phi:=\tr_E((\rho_E -\rho_E')\ln \rho_E)$ \cite{LandiReview2021} and we obtain
\begin{equation}
\label{Onsagerslike0}
\Sigma \geq F(\Phi, \langle \langle \ln \rho_E \rangle \rangle_t, \langle \langle \ln \rho_E \rangle \rangle_0),
\end{equation}
where we used $F(-x,y,z)=F(x,y,z)$, with $\langle \langle \ln \rho_E \rangle \rangle_{t}= \langle \langle \ln \rho_E \rangle \rangle_{\rho_E'}$ and $\langle \langle \ln \rho_E \rangle \rangle_{0}= \langle \langle \ln \rho_E \rangle \rangle_{\rho_E}$. It yields the explicit expression from (\ref{form12}),
\begin{equation}
\label{Onsagerslike1}
\Sigma \geq r\ln(r/s) + (1-r)\ln\big(\frac{1-r}{1-s}\big),
\end{equation}
for $r,s \in [0,1]$, where $r:=1/2+b/(4\Phi v)$, $s:=r-\Phi/2v$, $b:=\Phi^2+\langle \langle \ln \rho_E \rangle \rangle_0 - \langle \langle \ln \rho_E \rangle \rangle_t$, $v:=\sqrt{\langle \langle \ln \rho_E \rangle \rangle_t +b^2/(4\Phi^2)}$, which is the thermodynamic uncertainty relation for quantum entropy production with respect to the quantum entropy flux. The change $\Phi \rightarrow -\Phi$ leaves the bound invariant in (\ref{Onsagerslike1}), although swapping states $\rho_E' \leftrightarrow \rho_E$ does not. Note that if $\Phi=0$, then the relation (\ref{Onsagerslike1}) reduces to $\Sigma \geq 0$, which is simply the second law. Alternatively, for $\Phi \neq 0$, relation (\ref{Onsagerslike1}) establishes a positive lower bound for $\Sigma$.

{\bf \emph{Conclusions - }}
We introduced a relation (\ref{secondmain}) between the statistics of quantum observables (mean and variances) and $\chi_\lambda [\rho,\sigma]$, which is a generalization of the $\chi^2$ divergence for states $\rho$ and $\sigma$. We then used this relation to obtain a lower bound for the quantum relative entropy (\ref{form12}). When applied to the usual setup of quantum thermodynamics, this lower bound results in a thermodynamic uncertainty relation (TUR) for quantum entropy production (\ref{main}). For a particular case, the lower bound is expressed in terms of the entropy flux (\ref{Onsagerslike0}). Additionally, we showed that our result also reproduces a classic result from stochastic thermodynamics (\ref{discussion1}).

Furthermore, we provided a characterization of $\chi_\lambda [\rho,\sigma]$ in terms of operators (\ref{qchioperator}), obtained a TUR for the generalized triangular discrimination ($\lambda=1/2$), which is closely related to previous symmetric TURs in quantum systems, and derived the quantum Cramér-Rao inequality (\ref{unit5}) as a limiting case. We believe that our results (\ref{secondmain}) and (\ref{main}) are very general and may find applications in other problems not discussed in this letter, such as quantum speed limits and quantum computing.

{\bf \emph{Appendix- }}
We prove inequality (\ref{form3}) for complex random variables. Consider probabilities $P,Q$ in $s \in \mathcal{S}$, $\sum_s P(s)= \sum_s Q(s)=1$ and a complex valued random variable $x(s) \in \mathbb{C}$ with $Q(s)=0 \rightarrow P(s)=0$ (absolute continuity). Let $\overline{x}_p:=\langle x \rangle_p=\sum_s x(s) p(s)$, for $p \in \{P,Q\}$. We have from definition,
\begin{equation}
\label{app1}
|\overline{x}_P - \overline{x}_Q|^2 = |\sum_{s} (x(s)-c) (P(s)-Q(s))|^2,
\end{equation}
for any complex $c$. Using Cauchy–Schwarz inequality, we obtain
\begin{equation}
\label{app2}
|\sum_{s} (x(s)-c) (P(s)-Q(s))|^2 \leq \langle (|x-c|^2 \rangle_{Q}\langle (\frac{P-Q}{Q})^2\rangle_{Q}.
\end{equation}
Combining (\ref{app1}) and (\ref{app2}) for $c=\overline{x}_{Q}$, it yields 
\begin{equation}
\label{app3}
|\overline{x}_P - \overline{x}_Q|^2 \leq \langle |x-\overline{x}_{Q}|^2 \rangle_{Q}\langle (\frac{P-Q}{Q})^2\rangle_{Q}.
\end{equation}
which results in (\ref{form3}) after replacing $\chi^2(P|Q)=\langle (P-Q)^2/Q^2 \rangle_Q$ in (\ref{app3}).

\bibliography{librarytest}

\end{document}